\documentstyle[12pt]{article}

\begin{document}

\title{Computation of Kolmogorov's Constant in Magnetohydrodynamic Turbulence}
\author{M.\ K.\ Verma and J.\ K.\ Bhattacharjee  \\ 
{\em Department of Physics, Indian Institute of Technology,} \\ {\em Kanpur
208016, INDIA}}
\date{\today }
\maketitle

\begin{abstract}
In this paper we calculate Kolmogorov's constant for magnetohydrodynamic
turbulence to one loop order in perturbation theory using the direct
interaction approximation technique of Kraichnan. We have computed the
constants for various $E^u(k)/E^b(k)$, i.e., fluid to magnetic energy ratios
when the normalized cross helicity is zero. We find that $K$ increases from
1.47 to 4.12 as we go from fully fluid case $(E^b=0)$ to a situation when $%
E^u/E^b=0.5$, then it decreases to 3.55 in a fully magnetic limit $(E^u=0)$.
When $E^u/E^b=1$, we find that $K=3.43$.
\end{abstract}

{\flushleft PACS number: 47.65.+a, 47.25.cg, 52.35}

\pagebreak
Kolmogorov \cite{Kolm41} hypothesized that the energy spectrum $E(k)$ of
fluid turbulence in the inertial range (i.e., length scale intermediate
between the energy feeding and energy damping ranges) is a power law with a
spectral index of $-5/3$, i.e.,
\begin{equation}
\label{fluid}E(k)=K_{Ko}\Pi ^{2/3}k^{-5/3},
\end{equation}
where $K_{Ko}$ is an universal constant called Kolmogorov's constant, $k$ is
the wavenumber, and $\Pi $ is the nonlinear cascade of energy which is also
equal to the dissipation rate of the fluid.

The calculations based on Direct interaction approximations \cite
{Kraich:71,Leslie}, renormalization group technique \cite{RG},
self-consistent mode coupling \cite{mode} etc.\ show that $K_{Ko} \approx
1.5 $ in three dimensions (3D). However, in two dimensions (2D) Kraichnan
\cite{Kraich:71}, Olla \cite{Olla}, and Nandy and Bhattacharjee \cite{JKB}
show that $K_{Ko} \approx 6.4$ in the region where inverse cascade of energy
occurs. Experiments and numerical simulations \cite{exptsim} yield constants
approximately same as those predicted by the above calculations.

Magnetohydrodynamic (MHD) turbulence is more complex than the fluid
turbulence. In MHD there are velocity field ${\bf u}$ and magnetic field $%
{\bf B=B_0+b}$, where ${\bf B_0}$ is the mean magnetic field and {\bf b} is
the fluctuation in the magnetic field. It is customary to use Els\"asser
variables ${\bf z^{\pm }=u\pm b/\sqrt{4\pi \rho }}$ that can be interpreted
as Alfv\'en waves having positive and negative velocity and magnetic field
correlations respectively. Here $\rho $ is the density of the magnetofluid.
In this paper we assume incompressible approximation which in most of the
situations correspond to a constant density of the fluid. Matthaeus and
Zhou, Zhou and Matthaeus, and Marsch \cite{Kolm} constructed a phenomenology
when $z^{\pm }\gg B_0/\sqrt{4\pi \rho }$, where $B_0$ is the mean magnetic
field or the magnetic field of the largest eddies. In this phenomenology,
the energy spectra $E^{\pm }(k)$ of $z^{\pm }$ fluctuations are proportional
to $k^{-5/3}$ and
\begin{equation}
\label{eq:mhd}E^{\pm }(k)=K^{\pm }(\Pi ^{\pm })^{4/3}(\Pi ^{\mp
})^{-2/3}k^{-5/3},
\end{equation}
where $K^{\pm }$ are constants, which we will refer to as Kolmogorov's
constants for MHD turbulence, and $\Pi ^{\pm }(k)$ are the nonlinear cascade
rates of $z^{\pm }$ fluctuations respectively. The normalized cross helicity
$\sigma _c$, (strictly speaking spectral normalized cross helicity) defined
as $(E^{+}(k)-E^{-}(k))/(E^{+}(k)+E^{-}(k))$, play an important role in MHD
turbulence. In this paper, however, we restrict ourselves to zero normalized
cross helicity where, by symmetry, $\Pi ^{+}=\Pi ^{-}$ and $K^{+}=K^{-}=K$.
This choice of $\sigma _c$ is motivated by the simplicity of the calculation
and by the fact that in the solar wind most of the fluctuations in the outer
heliosphere (beyond earth) and some of them in the inner heliosphere (from
sun to earth) have negligible $\sigma _c$. There is another important energy
called residual energy $E^R$ defined as the difference between the kinetic ($%
E^u$) and magnetic ($E^b$) energies. Note, however, that the residual energy
is not conserved while both $E^{+}$ and $E^{-}$ are conserved.

In the other limit when $z^{\pm} \ll B_{0}/\sqrt{4 \pi \rho}$, Kraichnan and
Dobrowolny et al.\ \cite{mhdkraich} argued that
\begin{equation}
\label{eq:Kraich}\Pi^{+} = \Pi^{-} = \frac{1}{A^{2}} E^{+}(k) E^{-}(k)
k^{3},
\end{equation}
where $A$ is a constant called Kraichnan's constant.

The solar wind observations indicate that the it exhibits MHD turbulence.
The energy spectra of the solar wind has been found to be closer to $%
k^{-5/3} $ than to $k^{-3/2}$ \cite{MG:82}. However, it should be noted that
these exponents are difficult to distinguish. Recently we have performed
direct numerical simulations of MHD turbulence \cite{Verma:sim} in that we
could not conclude whether the spectral index was 5/3 or 3/2. However, the
nonlinear energy cascade rates $\Pi^{\pm}$ appear to follow Eq.~(\ref{eq:mhd}%
) rather than Eq.~(\ref{eq:Kraich}). Motivated by the above mentioned solar
wind observations, in this paper we assume that the energy spectra $%
E^{\pm}(k)$ obey Kolmogorov-like phenomenology [Eq.~(\ref{eq:mhd})] for $%
z^{\pm} > B_{0}$.

We determine for the first time the constant $K$ associated with the
Kolmogorov spectrum in MHD turbulence. To this end we attempt to find $K$ of
Eq.\ (\ref{eq:mhd}) using perturbative technique similar to DIA (see
Fournier et al.\thinspace and Camargo and Tasso \cite{mhdrg} for application
of Renormalization Group techniques to MHD). In this paper we only discuss
the situations when $E^{+}(k)=E^{-}(k)$ and $B_0=0$. However, we vary $%
E^u/E^b$ from 0 to $\infty $, i.e., from magnetic case to fully fluid case.
The computation of $K^{\pm }$ when $E^{-}(k)/E^{+}(k)$ is different from
unity and when $B_0$ is nonzero is under progress; forthcoming results will
be presented elsewhere.

The MHD equations in terms of $z^{\pm }$, in absence of $B_0$, are given by
\cite{mhdkraich}
\begin{equation}
\label{mhdeqn}
\begin{array}{c}
\frac d{dt}z_i^{\pm }(
{\bf k},t)+\nu _{+}k^2z_i^{\pm }({\bf k},t)+\nu _{-}k^2z_i^{\mp }({\bf k}%
,t)=f_i^{\pm } \\ +\epsilon M_{ijm}({\bf k})\int d{\bf p}z_j^{\mp }({\bf p}%
,t)z_m^{\pm }({\bf k-p},t)
\end{array}
\end{equation}
where
\begin{equation}
M_{ijm}({\bf k})=k_jP_{im}({\bf k});\ P_{im}({\bf k})=\delta _{im}-\frac{%
k_ik_m}{k^2};\ \nu _{\pm }=\frac{\nu \pm c^2/(4\pi \sigma )}2,
\end{equation}
$\nu $ is kinematic viscosity, $c$ is the speed of light, $\sigma $ is
conductivity, $f^{\pm }$ are forcing, and $\epsilon $ is the expansion
parameter which is set to one at the end. We assume independent forcing of $%
z^{\pm }$, i.e., $\partial f^{s_1}/\partial f^{s_2}=\delta ^{s_1s_2}$, where
$s_1,s_2=\pm $. We solve for self-energy $\hat \Sigma $ (a $2\times 2$
matrix), which is defined as
\begin{equation}
\label{selfdef}\hat G=\hat G_0-\hat G_0\hat \Sigma \hat G,
\end{equation}
to first order using the technique of Wyld \cite{Wyld} and Leslie \cite
{Leslie} (the details of the calculation are be presented in \cite{diaPRE}).
In our calculation we postulate the frequency dependence of the $z-z$
correlations functions as \cite{Leslie}:
\begin{equation}
C^{\pm \pm }(k,\omega )=\frac 1{2\pi }\frac{C^{\pm }(k)}{(i\omega -\eta
_k^{\pm })};C^{\pm \mp }(k,\omega )=\frac 1{2\pi }\frac{C^R(k)}{(i\omega -%
\frac{\eta _k^{+}+\eta _k^{-}}2)},
\end{equation}
where $\eta _k^{+}$ and $\eta _k^{-}$ are the reciprocal of the mean
response time of $z_k^{+}$ and $z_k^{-}$ fluctuations respectively. The
self-energy can be written as a following $2\times 2$ matrix:
\begin{equation}
\label{sigmhd}\hat \Sigma =\left(
\begin{array}{cc}
\Sigma ^{++} & \Sigma ^{+-} \\
\Sigma ^{-+} & \Sigma ^{--}
\end{array}
\right) =\frac 12\left(
\begin{array}{cc}
\eta ^{+} & \eta ^{-} \\
\eta ^{+} & \eta ^{-}
\end{array}
\right) ,
\end{equation}
\begin{eqnarray}
\eta^{\pm}(k) & = & \lim_{\omega \rightarrow 0} \frac{1}{2} k^{2}
\int \int d\bf{p} d\omega'   \times \nonumber \\
& & [b_{1}(k,p,q) (G^{\mp \pm}({\bf p},\omega')+G^{\mp \mp}({\bf p},\omega'))
 	C^{R}({\bf k-p},\omega-\omega')  \nonumber \\
& & + b_{2}(k,p,q) (G^{\pm \mp}({\bf p},\omega')+G^{\pm \pm}({\bf p},\omega'))
 	C^{\mp}({\bf k-p},\omega-\omega') \nonumber \\
& & +b_{3}(k,p,q) (G^{\pm \pm}({\bf p},\omega')+G^{\pm \mp}({\bf p},\omega'))
 	C^{\mp}({\bf k-p},\omega-\omega') \nonumber \\
& & +b_{4}(k,p,q) (G^{\mp \pm}({\bf p},\omega')+G^{\mp \mp}({\bf p},\omega'))
 	C^{R}({\bf k-p},\omega-\omega')]
\label{eq:sigma}
\end{eqnarray}
where $G$ is the Green's function, $b_i(k,p,q)=k^{-2}B_i(k,p,q)$, and $B_i$%
's are the functions of $(x,y,z)$, the cosines of the angles between ({\bf %
p,q}), ({\bf q,k}), and ({\bf k,p}) respectively; $B_i$'s given in the
appendix of Leslie \cite{Leslie}. Substitution of $\hat \Sigma $ of (\ref
{sigmhd}) in Eq. (\ref{selfdef}) and ignoring kinematic viscosity and
resistivity as compared to $\hat \Sigma /k^2$ yields $\hat G$
$$
\hat G(k,\omega )=\frac 1{-\omega \left( \omega +i\frac{\eta ^{+}+\eta ^{-}}%
2\right) }\left(
\begin{array}{cc}
-i\omega +\frac{\eta ^{+}}2 & \frac{\eta ^{-}}2 \\ \frac{\eta ^{-}}2 &
-i\omega +\frac{\eta ^{+}}2
\end{array}
\right)
$$
Note that when $C^{+}(k)=C^{-}(k)=C^R(k)$, i.e., when the magnetic energy is
zero, the above equations yield $\eta ^{+}=\eta ^{-}=\eta _{fluid}$.

As mentioned earlier, in this paper we restrict ourselves to situations when
$C^{+}(k)=C^{-}(k)=C(k)$ and assume the Kolmogorov scaling law, i.e., $E(k)$
are proportional to $k^{-5/3}$. In this situation $\eta _k^{+}=\eta _k^{-}$
by symmetry. Since the energy spectra $E(k)$, which is equal to $C(k)/(4\pi
k^2)$, is proportional to $k^{-5/3}$, the correlation functions $C^{\pm }(k)$
and self-energies will be \cite{Leslie}
\begin{equation}
\label{eq:etak}C(k)=\frac 1{4\pi }K\Pi ^{2/3}k^{-11/3};\ \eta ^{\pm
}(k)=\Lambda \Pi ^{1/3}k^{2/3},
\end{equation}
where $\Lambda $ is a constant. We denote the ratio $C^R/C$ by $\alpha $.
Using the above quantities and with the change of variables $p=\zeta k$ and $%
q=\kappa k$ we obtain
\begin{eqnarray}
\frac{\Lambda^{2}}{K} & = & \frac{1}{4} \int \int d\zeta
d\kappa \zeta
\frac{\kappa^{-11/3}}{(\zeta^{2/3}+\kappa^{2/3})}  k^{-2}\\
\nonumber
& & \times
\left[
\alpha(b_{1}(k,p,q)+b_{4}(k,p,q))+b_{2}(k,p,q)+b_{3}(k,p,q) \right].
\label{eq:sigma}
\end{eqnarray}
The integrals of Eq.\ (\ref{eq:sigma}) suffer from the well known ``infrared
problem'' which comes from the strong dynamic coupling of fluctuations with
widely differing wavenumbers. Over the years, various techniques have been
developed to tackle the difficulty in the context of pure fluid turbulence.
The earlier methods (essentially introduction of a cut off) are discussed in
Leslie\cite{Leslie}. Later work involving either Renormalization group
technique \cite{RG}, or the Lagrangian or semi-Lagrangian pictures \cite
{Lagrange}, or self-consistent screening \cite{mode} shows how the theory
can be made naturally finite and cut off independent. The infrarred
difficulties associated with Eq.~(\ref{eq:sigma}) can be similarly avoided.
Knowing that the full theory is constrained to be finite, we adopt the
practical procedure of evaluating the integral in Eq.~(\ref{eq:sigma}) with
a cut off $k_0=\lambda k$ and choosing $\lambda =1$ so that the pure fluid
value of $\Lambda ^2/K$ are obtained correctly when $\alpha =1$. Thereafter $%
\lambda $ is not varied. With this procedure we obtain $\Lambda ^2/K$ as a
function of $\alpha $. For some of the characteristic values of $\alpha $, $%
\Lambda ^2/K$ is listed in Table 1.

To obtain the numerical value of $K$ we need another equation involving $%
\Lambda $ and $K$. To this end, we derive an expression for energy cascade
rate $\Pi $ in terms of correlations $C(k)$ and Green's functions $G(k)$.
Following an approach similar to that of Leslie \cite{Leslie}, when $%
C^{+}(k)=C^{-}(k)=C(k)$, we obtain
\begin{equation}
\left( \frac d{dt}+2\nu _{+}k^2\right) E(k)+2\nu _{-}k^2E^R(k)=T(k,t).
\end{equation}
Using $T(k,t)$ we can write an expression for energy cascade rate, which is $%
\Pi =-\int_0^kdk^{\prime }T(k^{\prime },t)$. Substitution of $C$'s and $\eta
$'s, and change of variables $k^{\prime }=k/u,p=v(k/u)$, and $q=w(k/u)$
yield an expression for $\Lambda /K^2$, that is,
\begin{equation}
\frac \Lambda {K^2}=\int_0^1dv\ln \frac 1v\int_{v^{*}}^{1+v}dw[\Psi
(1,v,w)+\Psi (1,w,v)]
\end{equation}
where\
\begin{eqnarray}
\Psi(1,v,w)& = &
\frac{1}{4} v w (1+v^{2/3}+w^{2/3})^{-1}  \\
\nonumber
& & \times [
b_{1}(1,v,w) w^{-11/3} ( (1+\alpha^{2}) v^{-11/3}-2
\alpha) \\ \nonumber
&
& \mbox{} + (1+\alpha^{2}) b_{2}(1,w,v) v^{-11/3} (w^{-11/3}-1) \\
\nonumber
& & \mbox{} - 2 \alpha
b_{3}(1,v,w) w^{-11/3} - (1+\alpha^{2})
b_{4}(1,v,w) v^{-11/3} ],
\end{eqnarray}
and $v^{*}={\rm max}(v,|1-v|)$. See Table 1 for $\Lambda /K^2$ for various
values of $\alpha $.

We calculate $K$ from $\Lambda ^2/K$ and $\Lambda /K^2$. The values of $K$
for various values of $\alpha $ are listed in Table 1. We have also
calculated the constant $K$ in 2D. In 2D when $E^u=E^b$, we choose $\lambda
=1$, same as $\lambda $ of 3D, but not 0.065 which yields $K=6.6$ in fluid
limit. The choice of same $\lambda $ in 2D MHD was motivated by fact that in
MHD turbulence, forward cascade of energy and inverse cascade of cross
helicity occur in both 2D and 3D\cite{Ting}. Note that the behaviour of
fluids turbulence in 2D and in 3D are dramatically different.

The results presented in this paper are in good agreement with the
simulation results of Verma et al.\ \cite{Verma:sim} ($K=3.7$ in
three-dimensions for $E^{-}/E^{+}=0.6$, and average $K=6.6$ in
two-dimensions for $E^{-}/E^{+}=1,0.6$). In 3D simulation, however, only a
single run with a relatively lower resolution ($128^3$) was performed. We
need to perform more runs to come to a definite conclusion.

We find that $K$ monotonically increases from 1.47 to 4.12 as we go from
fully fluid case ($E^R=E$) to a situation when $E^R/E=-1/3$, i.e., $E^u/E^{{b%
}}=0.5$, then it decreases and finally reaches 3.55 in fully magnetic limit (%
$E^R=-E$). The maxima of $K$ occurring for $E^u/E^b=0.5$ is a curious result
because average $E^u/E^b$ for the solar wind in the outer heliosphere, where
$E^{+}\approx E^{-}$ is a good approximation, is 0.5. Note that maximum $K$
corresponds to minimum energy dissipation, hence it appears as if the solar
wind is settling to a minimum dissipation state. Further investigation in
this direction could possibly yield interesting results regarding minimum
dissipation states.

Recent temperature evolution studies of solar wind plasma show that
turbulent heating in the solar wind can account for the observed temperature
evolution when $K=1.0$ for nonAlfv\'{e}nic streams (streams for which $E^{+}
\approx E^{-}$) \cite{Verma:temp}. In the inner heliosphere, where the solar
wind is fluid dominated, the turbulent heating is possibly the major heating
mechanism. However, in the outer heliosphere, since $K$ is in the range of 3
to 4, turbulent heating could account for only 20-25 \% of the total
dissipation; rest of the dissipation should be provided by other sources,
e.g., shock heating, stream-stream interactions etc.

To conclude, we have calculated the Kolmogorov's constant for MHD turbulence
for zero $\sigma _c$ to one loop order using perturbative technique similar
to the direct interaction approximation of Kraichnan. We find that in fluid
dominated case, $K\approx 1.5-2$, but as magnetic energy increases, $K$
increases as well and reaches a maximum value of 4.12 when $E^b=2E^u$, then
it decreases and reaches 3.55 for a fully magnetic case. These results are
consistent with the recently performed simulation results and also shed
light on the evolution of $E^u/E^b$ ratio and temperature in the solar wind.

We thank Malay Nandy for fruitful discussion and R. Sethuraman for help in
computation. We also thank one of the referees for careful reading of the
manuscript and for useful comments.

\pagebreak

\newpage

\begin{table}
\begin{center}
\begin{tabular}{|c|c|c|c|c|} \hline
$E^{R}/E$ & $%
E^{u}/E^{b}$ & $\Lambda^{2}/K$ &
$\Lambda/K^{2}$ & $K$ \\
\hline
1 & $%
\infty$   & 0.114 & 0.190 & 1.47 \\
\hline
0      & 1     & 0.176 & 0.066
& 3.43 \\ \hline
$-1/3$ & $1/2$ &
0.197 & 0.053 & 4.12 \\ \hline
$-2/3$ &
$1/5$ & 0.218 & 0.056 & 4.11 \\
\hline
$-1$   & 0     & 0.239 & 0.073 &
3.55 \\ \hline \hline
$0^{2D}$ & 1
& 0.070 & 0.021 & 5.41 \\ \hline
\end{tabular}
\end{center}
\caption{Kolmogorov's constant $K$ of MHD turbulence for various values
of
$\alpha = E^{R}/E$ (assuming $E^{+}=E^{-}$). The symbols $E, E^{R},
E^{u}, E^{b}$
denote the total
energy, residual energy, kinetic and
magnetic energy
respectively.}

\end{table}

\end{document}